\documentclass{jpsj2}

\def\runtitle{Stochastic transition between turbulent branch and 
thermodynamic branch of an inhomogeneous plasma}
\def\runauthor{Mitsuhiro \textsc{Kawasaki}, Sanae-I. \textsc{Itoh}, 
Masatoshi \textsc{Yagi} and Kimitaka \textsc{Itoh}}

\title{%
Stochastic Transition between Turbulent Branch and \\
Thermodynamic Branch of an Inhomogeneous Plasma
}

\author{%
Mitsuhiro \textsc{Kawasaki}\thanks{mituhiro@riam.kyushu-u.ac.jp}, 
Sanae-I. \textsc{Itoh}, Masatoshi \textsc{Yagi} and 
Kimitaka \textsc{Itoh}$^{1}$
}

\inst{%
Research Institute for Applied Mechanics, Kyushu University, Kasuga 816-8580\\
$^1$National Institute for Fusion Science, Toki 509-5292
}

\recdate{\today}

\abst{%
Transition phenomena between thermodynamic branch and turbulent
branch in submarginal turbulent plasma are analyzed with
statistical theory. Time-development of turbulent fluctuation is
obtained by numerical simulations of Langevin equation which
contains submarginal characteristics. Probability density functions and 
transition rates between two
states are analyzed. Transition from turbulent branch to thermodynamic 
branch occurs in almost entire region between subcritical bifurcation point 
and linear stability boundary.
}

\kword{%
plasma turbulence, submarginal turbulence, subcritical bifurcation, 
Langevin equation, transition rate, probability density function
}

\begin{document}
\sloppy
\maketitle

\section{Introduction}

There have been observed various kinds of formations and destructions of 
transport barriers. Both in edge and internal regions of high temperature 
plasmas, the dynamical change often occurs on the short time scale, 
sometimes triggered by subcritical bifurcation. These features naturally lead 
to the concept of transition \cite{Itoh95}. 

The transition takes place as a statistical process in the presence of 
statistical noise source which is induced by strong turbulence fluctuation. 
As a generic feature, the transition occurs with a finite probability when a 
controlling parameter approaches the critical value. 

The nonequilibrium statistical mechanics, which deals with dynamical phase 
transitions and critical phenomena, should be extended for inhomogeneous 
plasma turbulence \cite{Kubo85}. To this end, statistical theory for plasma 
turbulence has been developed and stochastic equations of motion 
(the Langevin equations) of turbulent plasma were derived \cite{Itoh99}. 
The framework to calculate the probability density function (PDF), 
the transition rates etc. have also been made.

In this paper, we apply the theoretical algorithm to an inhomogeneous
plasma with the pressure gradient and the shear of the magnetic
field. Microturbulence in the system is known to be subcritically excited 
from the thermodynamic branch \cite{Itoh96}. 
The transition between the thermodynamic branch 
and the turbulent branch is studied. We show that the transition occurs
stochastically by numerically solving the Langevin equation of the
turbulent plasmas. In order to characterize the stochastic nature of the
transition, the frequency of occurrence of a transition per unit time
(the transition rate) is calculated as a function of the
pressure-gradient and the plasma temperature. The results show that the
transition from the turbulent branch to the thermodynamic branch occurs in
a wide region instead of at a transition point.

\section{Theoretical Framework}

In this section, we briefly review the theoretical framework \cite{Itoh99} 
used in our analysis of turbulent plasmas. 

The theory is based on the Langevin equation, eq.~(\ref{Langevin-k}), derived 
by renormalizing with the direct-interaction approximation the reduced MHD 
for the three fields: the electro-static potential, 
the current and the pressure. 
The Langevin equation gives the time-development of the fluctuating parts of 
the three fields as 
\begin{equation}
\frac{\partial \mathbf{f}}{\partial t}+\hat{\cal L}\mathbf{f}=\mathbf{\cal N}
(t), \ \mbox{where} \ \mathbf{f}(t) \equiv 
\left( \begin{array}{c} \phi(t) \\ J(t) \\ p(t) \end{array} \right).
\label{Langevin-k}
\end{equation}
In this equation, the nonlinear terms are divided into two parts: One part is 
coherent with the test field $\mathbf{f}(t)$ and is included into the 
renormalized operator $\hat{\cal L}$. The other is incoherent and is modeled 
by a random noise $\mathbf{\cal N}(t)$.
Since $\mathbf{\cal N}(t)$ is a force which fluctuates randomly in time, 
the Langevin equation describes the stochastic time-development of the 
fluctuation of the three fields. 

By analyzing the Langevin equation eq.~(\ref{Langevin-k}), a number of 
statistical properties of turbulent plasmas can be derived. 
For example, the analytical formulae for the change rate of plasma states, 
the transition rates, were derived. Furthermore, since the renormalized 
transport coefficients come from the term of the random force 
$\mathbf{\cal N}(t)$, relations between the fluctuation levels of turbulence 
and the transport coefficients like the viscosity and the thermal diffusivity 
were derived.

\section{A Model}

With the theoretical framework briefly described in the previous section, 
we analyze a model of inhomogeneous plasmas with the pressure-gradient and 
the shear of the magnetic field \cite{Itoh99}. The model is formulated with 
the reduced MHD for the three fields of the electro-static potential, 
the current and the pressure. The shear of the magnetic field is given as 
$\mathbf{B}=(0,B_0 s x, B_0)$ where $B_0(x) = \mbox{const}\times (1+\Omega' 
x+\cdots)$. The pressure is assumed to change in $x-$direction.

\begin{figure}[hptd]
 \begin{center}
 \includegraphics[width=8cm, height=5cm, keepaspectratio]{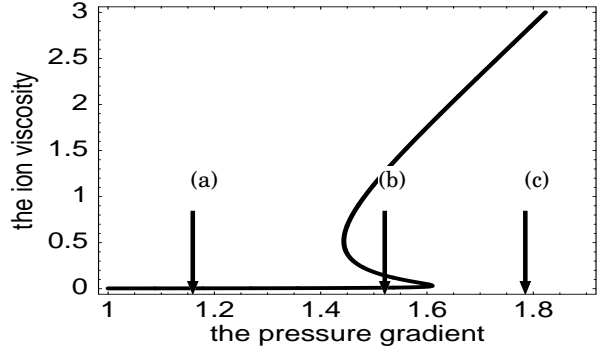}
 \end{center}
 \caption{The dependence of the renormalized ion-viscosity on 
the pressure-gradient. It is clearly seen that the bifurcation between a low 
viscosity state (the thermodynamic branch) and a high viscosity state 
(the turbulent branch) occurs. 
The arrows indicate the typical values of the pressure-gradient where some 
physical quantities are evaluated in the rest of this paper.}
 \label{ion-viscosity}
\end{figure}

It has been known that in this system bifurcation due to the subcritical 
excitation of the current diffusive interchange mode (CDIM) occurs 
\cite{Itoh96} as shown in Fig.\ \ref{ion-viscosity}.

Figure\ \ref{ion-viscosity} shows the dependence of
the turbulent ion-viscosity on the pressure-gradient. 
The turbulent ion-viscosity is proportional to the fluctuation level. 
Both the pressure-gradient and the turbulent viscosity are normalized. 
It is clearly seen that the bifurcation between a low viscosity
state and a high viscosity state occurs. Due to the bifurcation,
transition between the two states and hysteresis are expected to be
observed. The low viscosity state is called ``the thermodynamic branch'',
since this state is continually linked to thermal equilibrium. 
We call the high viscosity state ``the
turbulent branch'', since the fluctuation level is also large in a strong
turbulent limit \cite{Itoh99}. The ridge point where the turbulent
branch ends is denoted ``the subcritical bifurcation point''. The region
between the subcritical bifurcation point and the ridge near the linear
stability boundary is called ``the bi-stable regime''.

From the deterministic point of view, the transition from the thermodynamic 
branch to the turbulent branch is expected to occur {\it only} at the
ridge point near the linear stability boundary and the transition in the
opposite direction is expected to occur {\it only} at the subcritical 
bifurcation point.

\section{Stochastic Occurrence of the Transition}
\begin{figure}[hptd]
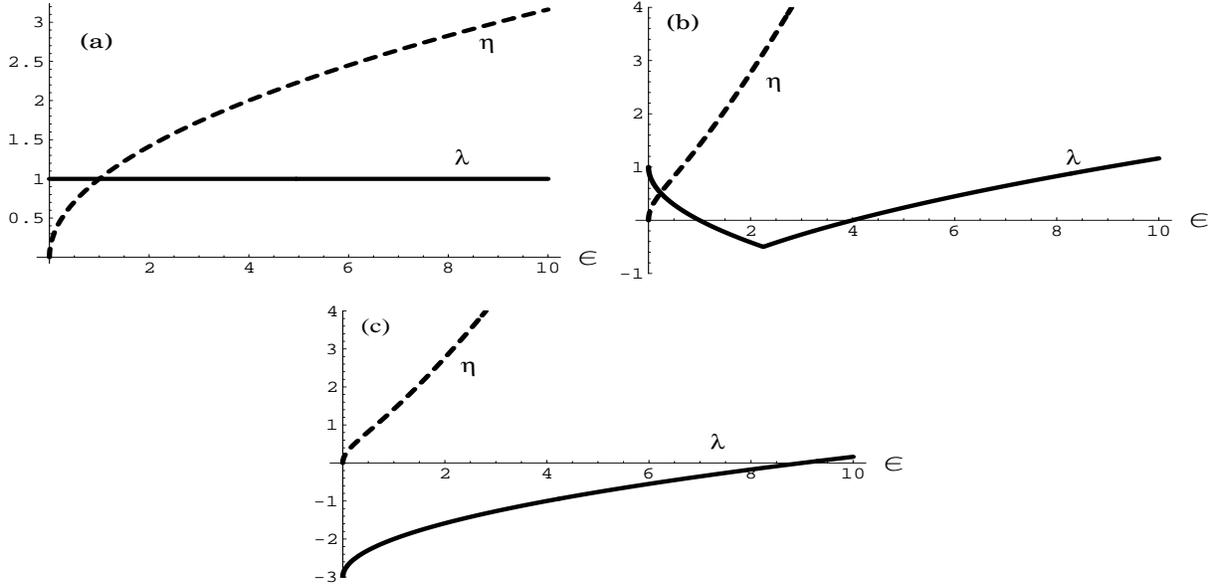

\begin{center}
 \includegraphics[width=8cm, height=4cm]{fig2a.eps}
 \includegraphics[width=8cm, height=4cm]{fig2b.eps} 
 \includegraphics[width=8cm, height=4cm]{fig2c.eps}
\end{center}
 \caption{(a) The coefficients $\lambda(\varepsilon)$ and $\eta(\varepsilon)$ 
in eq.~(\ref{Langevin}) at (a) in Fig.\ \ref{ion-viscosity}. 
(b) $\lambda(\varepsilon)$ and $\eta(\varepsilon)$ at (b) in Fig.\ 
\ref{ion-viscosity}.
(c) $\lambda(\varepsilon)$ and $\eta(\varepsilon)$ at (c) in Fig.\ 
\ref{ion-viscosity}.}
\label{lambda-eta}
\end{figure}
\begin{figure}[hptd]
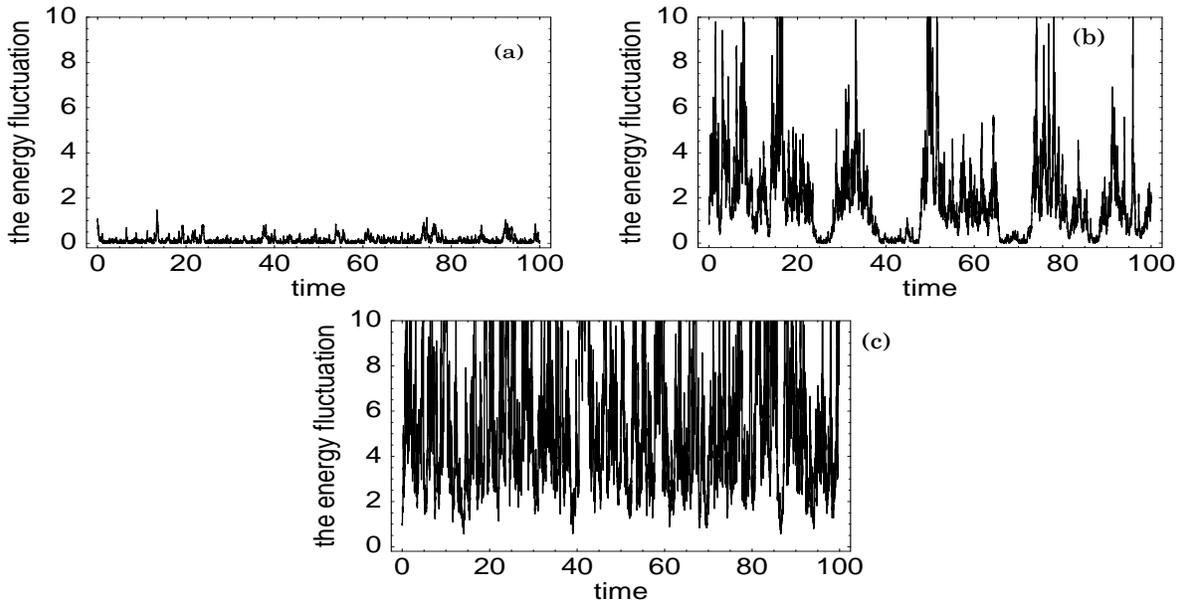

 \begin{center}
 \includegraphics[width=8cm, height=4cm]{fig3a.eps}
 \includegraphics[width=8cm, height=4cm]{fig3b.eps}
 \includegraphics[width=8cm, height=4cm]{fig3c.eps}
 \end{center}
 \caption{(a) A sample of a time-series of $\varepsilon(t)$ for the point 
(a) in Fig.\ \ref{ion-viscosity}. 
There is only small fluctuation since the system is always in the 
thermodynamic branch. 
(b) $\varepsilon(t)$ for the point (b) in Fig.\ \ref{ion-viscosity}. 
Bursts are observed intermittently. 
It means that the transition between the thermodynamic branch and the 
turbulent branch occurs stochastically. 
(c) $\varepsilon(t)$ for the point (c) in Fig.\ \ref{ion-viscosity}. 
Bursts occur simultaneously since the system in the turbulent branch.}
 \label{time-series}
\end{figure}
In order to capture the characteristics of the two states, we concentrate on 
the time-development of the energy of fluctuation of the electric field, 
$\varepsilon(t)$. The quantity $\varepsilon(t)$ obeys the 
coarse-grained Langevin equation, eq.~(\ref{Langevin}), which has been derived 
in ref. 3.
\begin{equation}
\frac{d}{dt}\varepsilon(t)=-2 \lambda(\varepsilon)\varepsilon(t)+
\eta(\varepsilon)R(t).
\label{Langevin}
\end{equation}
Here, $R(t)$ is assumed to be the Gaussian noise whose variance is unity. 
The coefficients 
$\lambda(\varepsilon)$ and $\eta(\varepsilon)$ depend on the 
pressure-gradient and the temperature and so the shapes of the functions 
for one regime of the pressure-gradient 
are completely different from that for the other regime. 
The shapes of $\lambda(\varepsilon)$ and $\eta(\varepsilon)$ for each of 
three regimes are shown in Fig.\ \ref{lambda-eta}. 
For the detailed formulae of $\lambda(\varepsilon)$ and $\eta(\varepsilon)$, 
see ref. 5.

The essential point is that the function 
$\lambda(\varepsilon)$ takes both a positive and a negative value in the 
bi-stable regime (Fig.\ \ref{lambda-eta} (b)). 
So, the fluctuation of the electric field is suppressed 
when $\lambda(\varepsilon)$ is positive and it is excited when 
$\lambda(\varepsilon)$ is negative. 
Consequently, there are two metastable states in the bi-stable regime. 

By solving numerically eq.~(\ref{Langevin}), we obtain the following samples 
of a time-series for each of three values of the pressure-gradient 
(three points (a), (b) and (c) shown in Fig.\ \ref{ion-viscosity}). 
When the pressure-gradient is fixed at the value smaller than the subcritical 
bifurcation value ((a) in Fig.\ \ref{ion-viscosity}),  
there is only small fluctuation since the system is always in the 
thermodynamic branch as shown in Fig.\ \ref{time-series} (a). 

On the other hand, when the pressure-gradient takes a value in the bi-stable 
regime ((b) in Fig.\ \ref{ion-viscosity}), 
bursts are observed intermittently as shown in Fig.\ \ref{time-series} (b). 
That is, transition between the thermodynamic branch and the turbulent branch 
occurs {\it stochastically}. The bursts correspond to the turbulent branch 
and the laminar corresponds to the thermodynamic branch. The fact that the 
residence times at the each states are random leads to the statistical 
description of the transition with the transition rates described in the 
rest of this paper. 

When the value of the pressure-gradient is larger than that of the linear 
stability boundary ((c) in Fig.\ \ref{ion-viscosity}), 
bursts are always observed (see Fig.\ \ref{time-series} (c)). 
It means that the system is always in the turbulent branch.

\section{The Probability Density Functions}
In order to characterize the random fluctuation shown in the last section, 
we introduce the probability density function (PDF) $P(\varepsilon)$ 
defined as the probability density that a random variable takes 
a certain value $\varepsilon$. 
PDFs often reveal the invisible structures hidden in randomly fluctuating data 
$\varepsilon(t)$. 

By counting the frequency of realization of a certain value $\varepsilon$ 
from time-serieses of $\varepsilon(t)$ over sufficiently long time, 
the histogram is obtained. The PDF $P(\varepsilon)$ is the histogram 
normalized with the total frequency. So, in general, the PDF can be obtained 
from time-serieses observed in real experiments as a normalized histogram.

\begin{figure}[hptd]
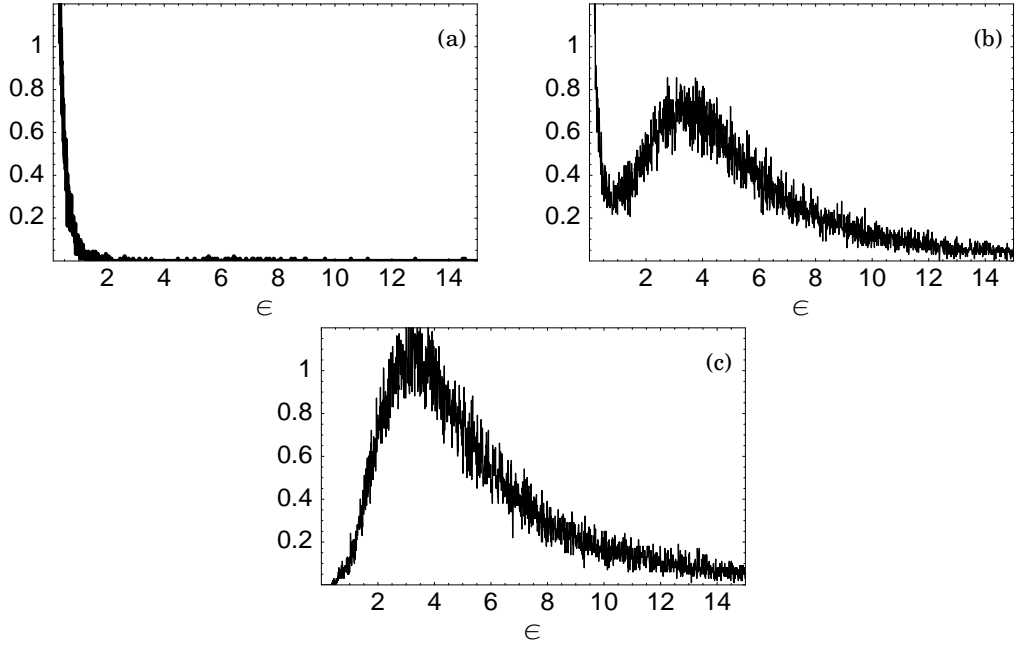

 \begin{center}
 \includegraphics[width=7cm, height=5cm, keepaspectratio]{fig4a.eps}
 \includegraphics[width=7cm, height=5cm, keepaspectratio]{fig4b.eps}
 \includegraphics[width=7cm, height=5cm, keepaspectratio]{fig4c.eps}
 \end{center}
 \caption{(a) The PDF when the value of the pressure-gradient is smaller than 
that of the subcritical bifurcation point 
((a) in Fig.\ \ref{ion-viscosity}). 
The PDF takes a finite value only in the small $\varepsilon$ region. 
(b) The PDF when the value of the pressure-gradient is fixed in the 
bi-stable regime ((b) in Fig.\ \ref{ion-viscosity}). 
The two peaks are clearly seen, even though 
it is not clear from the time-series Fig.\ \ref{time-series} (b) 
that there are two regions where the system is found frequently. 
(c) The PDF when the value of the pressure-gradient is larger than that 
of the linear stability boundary ((c) in Fig.\ \ref{ion-viscosity}). 
There is single peak corresponding to 
the turbulent branch.}
 \label{pdf}
\end{figure}

Figure\ \ref{pdf} shows the PDFs for three regimes of the 
pressure-gradient obtained from the time-series shown in 
Fig.\ \ref{time-series}.
Figure\ \ref{time-series} (b) shows that transition frequently occurs between 
the turbulent branch and the thermodynamic branch. 
Dithering between these two states is observed also in the PDF as 
the two peaks in Fig.\ \ref{pdf} (b). 

\begin{figure}[hptd]
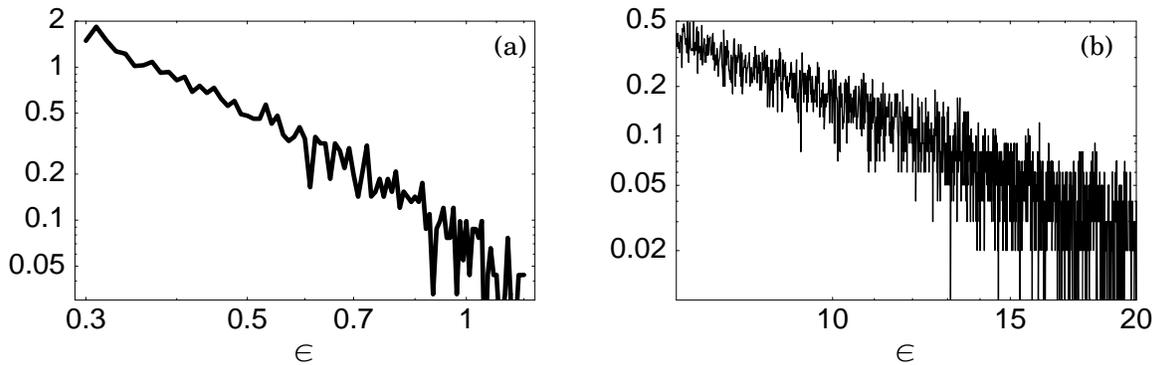

 \includegraphics[width=8cm, height=5cm, keepaspectratio]{fig5a.eps}
 \includegraphics[width=8cm, height=5cm, keepaspectratio]{fig5b.eps}
 \caption{(a) The tail of the PDF when the pressure-gradient is fixed at the 
value smaller than the subcritical bifurcation value. 
The PDF obeys the power-law for relatively large $\varepsilon$, 
even though the system is in the thermodynamic branch.
(b) The tail of the PDF when the value of the pressure-gradient is 
larger than that of the linear stability boundary. The PDF obeys the 
power-law when the system is in the turbulent branch.}
 \label{pdf-tail}
\end{figure}

Figure\ \ref{pdf-tail} (a) shows the tail of the PDF when the 
pressure-gradient is fixed at the value smaller than the subcritical 
bifurcation value. 
The PDF obeys the power-law for relatively large $\varepsilon$, 
even though the system is in the thermodynamic branch. 
The power-law tail when the value of the pressure-gradient is larger than 
that of the linear stability boundary is shown in Fig.\ \ref{pdf-tail} (b).

It is important from the point of view of theoretical investigation to note 
that the PDFs are also obtained as solutions of the Fokker-Planck equation 
equivalent to the Langevin equation eq.~(\ref{Langevin}).

\section{The Transition Rates}
In order to formulate the stochastic transition phenomena in the
bi-stable regime, we introduce the transition rates. There are
transitions in two opposite direction: the transition from the thermodynamic 
branch to the turbulent branch, which we call ``the forward
transition, and the transition in the opposite direction is called ``the
backward transition''. There are two transition rates. One is the forward 
transition rate $r_f$ which is the frequency of occurrence of the forward 
transition per unit time and the other is the backward transition rate $r_b$ 
defined similarly as the frequency of occurrence of the backward transition 
per unit time. 

It is important to note that these quantities are observable quantities. 
It is easily shown that the forward transition rate is equal to the average 
of inverse of the residence time at the thermodynamic branch and the backward 
transition rate is equal to the average of inverse of the residence time at 
the turbulent branch. Therefore, these transition rates can be measured from 
the time-serieses of fluctuation. 

\begin{figure}[hptd]
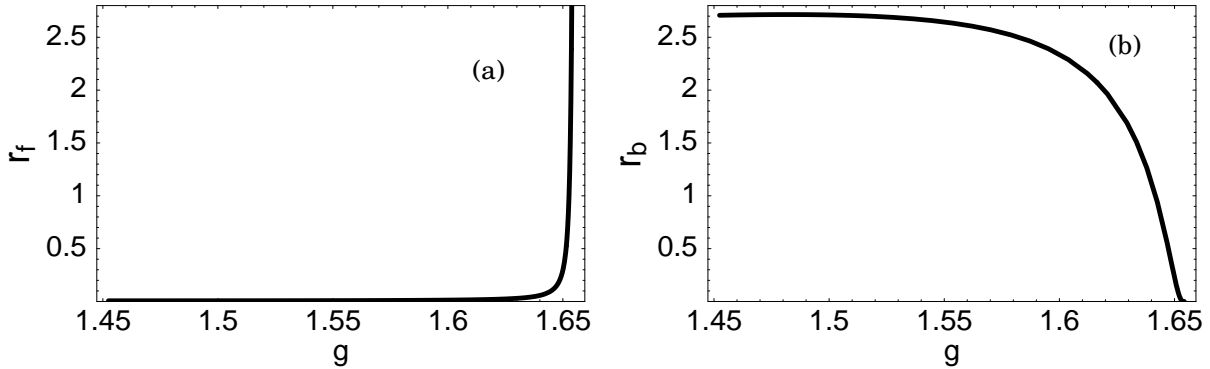

 \includegraphics[width=8cm, height=5cm, keepaspectratio]{fig6a.eps}
 \includegraphics[width=8cm, height=5cm, keepaspectratio]{fig6b.eps}
 \caption{(a) The dependence of the forward transition rate per unit time 
on the normalized pressure-gradient, $g$, in the bi-stable regime. 
The left edge and the right edge of the horizontal axis corresponds to the 
subcritical bifurcation point and the linear stability boundary. 
It is seen that the forward transition occurs mainly in the vicinity of the 
linear stability boundary. (b) The dependence of the 
backward transition rate per unit time on the pressure gradient 
in the bi-stable regime. It is seen that the backward transition occurs 
in the almost entire bi-stable regime.}
 \label{tr}
\end{figure}

We determine the value of the pressure-gradient with which the
transition occurs frequently. The transition rates are calculated with
the formulae derived in ref. 6. 
Figure\ \ref{tr} shows the dependence of 
the forward transition rate and the backward transition rate 
on the pressure gradient in the bi-stable 
regime, i.e., (b) in Fig.\ \ref{ion-viscosity}.

The forward transition triggered by the thermal noise occurs mainly in
the vicinity of the linear stability boundary. In contrast, it
is clearly seen that the backward transition occurs in the almost entire
bi-stable regime in contrast to the expectation from the deterministic point 
of view of bifurcation phenomena. This behavior is due to strong turbulent
fluctuation. It is noted that the backward transition, i.e. the
transition in a turbulence, occurs in a ``region'' instead of a ``point'' 
of the parameter space.

\section{Hysteresis Phenomena}
Up to now, the value of the pressure-gradient is fixed. Next, 
in order to investigate hysteresis phenomena, we turn to 
the case when the pressure-gradient changes in time.

It is important to investigate this case, since the pressure-gradient can be a 
dynamical variable in realistic plasmas. 
Since the characteristic time-scale of the transition is given by 
the inverse of the transition rate, $1/r$, the effect of the change of the 
pressure-gradient is governed by the interrelation between the transition 
rate $r$ and the time-rate of the change of the pressure-gradient, 
$\dot{g}(t)$. When $|\dot{g}(t)/g(t)| \sim r$ or $|\dot{g}(t)/g(t)| > r$, 
the system cannot follow the change of the pressure-gradient $g(t)$. 
Then, the state of the system depends on the value of the 
pressure-gradient $g(t)$ in the past and hysteresis phenomena are 
expected to be observed. 

On the other hand, when $|\dot{g}(t)/g(t)| \ll r$, 
the system can follow the change of the pressure-gradient $g(t)$. 
Since in this case the state of the system is the steady state for the 
value of the pressure-gradient at the moment, hysteresis phenomena cannot be 
observed.

\begin{figure}[hptd]
 \begin{center}
 \includegraphics[height=5cm, width=8cm, keepaspectratio]{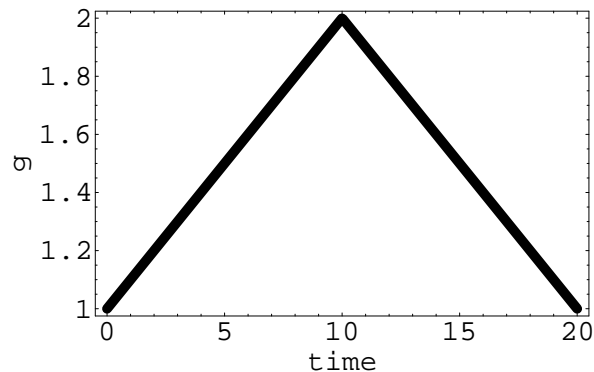}
 \end{center}
 \caption{The protocol to change the pressure-gradient to analyze 
hysteresis phenomena. 
At first the pressure-gradient is increased through the bi-stable regime 
($1.45<g<1.65$)and after that it is decreased to the original value ($g=1$). 
The time-rate of change of the pressure-gradient is assumed to be constant 
$|\dot{g}(t)|=0.1$ for simplicity.}
 \label{protocol-g-change}
\end{figure}

The protocol to change the pressure-gradient is as follows 
(see Fig.\ \ref{protocol-g-change}.): 
At first the pressure-gradient is increased through the bi-stable regime 
and after that it is decreased to the original value. 
The time-rate of change of the pressure-gradient is assumed to be constant 
for simplicity. Furthermore, in order to observe hysteresis, 
the speed of change of the pressure-gradient is chosen so that 
it is of the same order of the transition rates or larger than that. 

\begin{figure}[hptd]
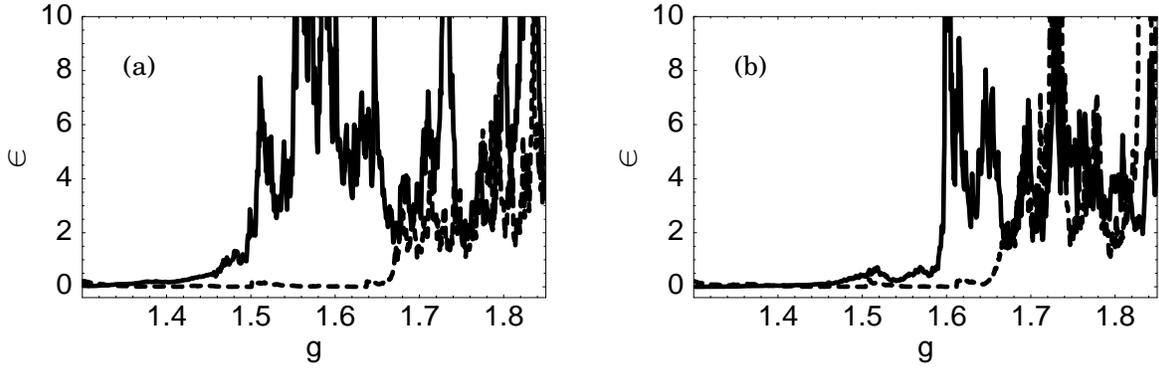

 \includegraphics[width=8cm, height=5cm, keepaspectratio]{fig8a.eps}
 \includegraphics[width=8cm, height=5cm, keepaspectratio]{fig8b.eps}
 \caption{(a) A sample of the hysteresis loop 
drawn by numerically solving the Langevin equation eq.~(\ref{Langevin}). 
The dashed line show $\varepsilon$ when the pressure-gradient is increased 
and the solid line show $\varepsilon$ when the pressure-gradient is 
decreased. (b) The other sample of the hysteresis loop obtained for the 
{\it exactly same} temperature and 
the time-rate of change of the pressure-gradient for Fig.\ \ref{hysteresis} 
(a). 
Since the backward transition can occur in the almost entire bi-stable 
regime, the backward transition point is different from that of 
Fig.\ \ref{hysteresis} (a).}
 \label{hysteresis}
\end{figure}

Figure\ \ref{hysteresis} (a) and (b) are the samples of 
hysteresis loops drawn by numerically solving the Langevin equation 
eq.~(\ref{Langevin}). 
The dashed lines show $\varepsilon$ when the pressure-gradient is increased 
and the solid lines show $\varepsilon$ when the pressure-gradient is decreased.
These figures, Fig.\ \ref{hysteresis} (a) and (b), 
are obtained for the {\it exactly same} temperature and 
the time-rate of change of the pressure-gradient. 
The forward transition occurs around at $g=1.65$ in both cases shown in 
Fig.\ \ref{hysteresis} (a) and (b). 
However, the backward transition point shown in Fig.\ \ref{hysteresis} (a) 
around at $g=1.5$ is completely different from the transition point around 
$g=1.6$ shown in Fig.\ \ref{hysteresis} (b). 
It is because the backward transition can occur in the almost 
entire bi-stable regime as shown in the analysis of 
the backward transition rate (see Fig.\ \ref{tr} (b)). 
Consequently, the backward transition point changes {\it stochastically} 
from case to case.
Since the distribution of the transition point is due to strong turbulent 
fluctuation, it is expected in general that transition points between 
different turbulent branches are distributed stochastically.
We expect that distribution of the L/H transition point 
observed in real experiments \cite{ITER94} is explained in this direction.

\section{Summary and Discussion}
Summarizing our work, we applied the statistical theory of plasma
turbulence to problems of the transition phenomena of submarginal
turbulence. By numerically solving the Langevin equation, the typical
time-development of fluctuation is obtained. It tells that the
transition for the model of inhomogeneous plasma occurs stochastically
and suggests how the transition phenomena due to subcritical bifurcation
may look in time-serieses obtained in real experiments. 

Furthermore, we obtained the PDFs of $\varepsilon(t)$ and 
the pressure-gradient dependence of the transition rates. It is
shown that the backward transition occurs with almost equal frequency in
the entire bi-stable regime, so the transition occurs in a
``region''. 
The concept ``transition region'' is necessary in the analysis of data 
obtained by real experiments.
It is confirmed that the backward transition does not occur only at the 
bifurcation point but occur also in the region of the 
pressure-gradient by observing the 
hysteresis loops obtained by numerically solving the Langevin equation.

It is important to discuss whether the transition phenomena 
considered in this paper can be observed in real experiments. 
Since the characteristic time-scale of the 
transition is given by the inverse of the transition rate, 
observability depends on the interrelation between the time resolution of 
observation $\triangle t$ and the transition rate $r$. 
When $\triangle t$ is much smaller than $1/r$, the transition phenomena 
as shown in Fig.\ \ref{time-series} (a) are expected to be observed. 
On the other hand, 
when $\triangle t$ is of the same order of $1/r$ or larger than $1/r$,  
transition phenomena average out and 
only the average over $\triangle t$ is observed.
This discussion is generic regardless of the type of transition, e.g. 
the transition between the thermodynamic branch and the turbulent branch, 
L/H transition etc.

\section*{Acknowledgments}
We wish to acknowledge valuable discussions with Atsushi Furuya. 
We thank Akihide Fujisawa for showing us unpublished 
experimental results which inspired us. 
The work is partly supported by 
the Grant-in-Aid for Scientific Research of Ministry of Education, 
Culture, Sports, Science and Technology, the collaboration programmes of 
RIAM of Kyushu University and the collaboration programmes of NIFS.

\end{document}